%% file: main.tex
\documentclass[conference]{IEEEtran}
\IEEEoverridecommandlockouts
\usepackage{cite}
\usepackage{amsmath,amssymb,amsfonts}
\usepackage{algorithmic}
\usepackage{graphicx}
\usepackage{textcomp}
\usepackage{amsthm}
\usepackage{mathtools}

\def\red{\color{red}}

\usepackage{tcolorbox}
\newtheorem{theorem}{Theorem}
\usepackage{hyperref}

\usepackage{xcolor}
\def\BibTeX{{\rm B\kern-.05em{\sc i\kern-.025em b}\kern-.08em
    T\kern-.1667em\lower.7ex\hbox{E}\kern-.125emX}}
    
\begin{document}

\title{
Goal-Oriented Communications for Remote Inference under Two-Way Delay with Memory\\
\thanks{This work was supported in part by TUBITAK grant 22AG019, NSF grant CNS-2239677, ARO grant W911NF-21-1-0244. Cagri Ari was also supported by Turk Telekom within the framework of 5G and Beyond Joint Graduate Support Programme coordinated by Information and Communication Technologies Authority.}
}

\author{\IEEEauthorblockN{Cagri Ari\textsuperscript{\dag}, Md Kamran Chowdhury Shisher\textsuperscript{\ddag}, Elif Uysal\textsuperscript{\dag}, and Yin Sun\textsuperscript{\ddag}}
\IEEEauthorblockA{\textit{Department of Electrical and Electronics Engineering, Middle East Technical University, Ankara, Turkiye}\textsuperscript{\dag} \\
\textit{Department of
Electrical and Computer Engineering, Auburn University, AL, USA}\textsuperscript{\ddag} \\
ari.cagri@metu.edu.tr, mzs0153@auburn.edu, uelif@metu.edu.tr, yzs0078@auburn.edu}
}

\newif\iflogvar

\logvartrue

\iflogvar
  \include{technical_report}
\else
  \include{ISIT_final}
\fi

\end{document}

%% file: technical_report.tex
\maketitle
\begin{abstract}
We study the design of a goal-oriented sampling and scheduling strategy through a channel with highly variable two-way random delay, which can exhibit memory (e.g., Delay and Disruption Tolerant Networks). The objective of the communication is to optimize the performance of remote inference, where an inference algorithm (e.g., a trained neural network) on the receiver side predicts a time-varying target signal using the data samples transmitted by a sensor. Previous formulations to this problem either assumed a channel with IID transmission delay, neglecting feedback delay or considered the monotonic relation that the performance only gets worse as the input information ages. We show how, with delayed feedback, one can effectively exploit the knowledge about delay memory through an index-based threshold policy. This policy minimizes the expected time-average inference error that can be monotone or non-monotone in age. The index function is expressed in terms of the Age of Information (AoI) on the receiver side and a parameter regarding the distribution of subsequent transmission delay, both of which can readily be tracked.
\end{abstract}

\begin{IEEEkeywords}
Goal-Oriented Communications, Markovian Delay, Two-way Delay, Remote Inference, Age of Information, Delay with Memory, Disruption Tolerant Networks (DTN).
\end{IEEEkeywords}

\section{Introduction}
Orders of magnitude enhancement is desired in terms of performance, coverage, capacity, and energy efficiency in 6G with respect to existing cellular networks. Such striking objectives accompany the demand for intelligent applications that run based on remotely collected data, sometimes involving non-terrestrial links. Based on the envisioned coexistence of terrestrial and non-terrestrial connections in 6G, it will not be unusual for network connections to be serviced through multiple alternative paths with highly variable delay. The relays in the network may force the utilization of a selected path for a particular application, and this path may be changed in time to manage the variety of flows in the network. Therefore, while developing efficient policies to optimize the performance of intelligent applications, considering their adaptability to significantly varying delay conditions, specifically the time dependence in the delay statistics, is crucial. \par
On the other hand, for the increasing numbers of these intelligent applications, including remote monitoring, control, and inference over networks, the scalability of the network architecture requires a paradigm shift in the communication system and protocol design. This paradigm shift is for networks to focus on the effective accomplishment of the sensing task at the destination side rather than be a vehicle to solve the transmission problem \cite{shannon1949mathematical} whose goal is to reliably transmit the data produced by a source. Solving the effective communication problem in remote intelligence with efficient use of network resources requires combining the data generation and transmission problems whereby data samples that are most significant to the computation at the destination side are delivered in a timely manner\cite{uysal2022semantic}. We will refer to this paradigm as \textit{Goal-oriented} communication. Recent activities in the communication and control communities illustrate the value of such an approach in reducing communication requirements for a given level of application performance \cite{voortman2023remote, sagduyu2023multi, merluzzi2022effective, talli2023semantic}. \par
A goal-oriented communication network discards the assumption that data arrives at the communication system interface as part of an exogenous process. This implies that the communication link can pick and choose which data packets to transmit to the source for effective computation at the destination based on the state of the network (e.g., the delay state). As we do not expect the link layer to operate jointly with the application layer for all possible applications, a surrogate metric that the link layer can use when making the decisions of which packets to choose and send is useful. In this paper, following the model in \cite{shisher2022does}, we use Age of Information (AoI) as an intermediate metric that can be tracked by the link, irrespective of the particular application. A mapping of application performance with respect to the AoI is then all that is needed for the link layer to operate in a goal-oriented manner. This paper builds on the work in the series of papers  \cite{shisher2022does, sun2019sampling, ornee2021sampling,shisher2021age} that first demonstrated the usefulness of AoI metric as a surrogate for effective communication for remote tracking or inference of a process. \par
AoI is an indicator of the freshness of the data at the destination of a data flow \cite{kaul2012real}. The AoI at time $t$, $\Delta(t)$, is given by the relation $\Delta(t)=t-U_{t}$, where $U_{t}$ is the generation time of the most recently delivered data packet. This formulation is perhaps most meaningful when the data packets contain status updates, such that the most recent update makes all the past ones obsolete, yet age and functions thereof have also served very well in capturing performance related to freshness in a variety of other applications, through optimizing nonlinear functions of age. Examples of the use of generalized functions of AoI include control system scenarios \cite{klugel2019aoi, soleymani2019stochastic}, remote estimation \cite{sun2019sampling, ornee2021sampling, pan2023sampling}, and remote inference \cite{shisher2021age, shisher2022does, shisher2023learning}. While a number of studies considered the analysis of age in queuing models, closest to the spirit of this paper is the control of age via replacement of exogenous data arrivals with the generation of data "at will" \cite{bacinoglu2015age, sun2017update, kadota2018scheduling, kadota2019scheduling, sun2019sampling2, tripathi2019whittle, klugel2019aoi, sun2019closed, bedewy2021optimal}. A generalization of this approach is to incorporate jointly optimal sampling and scheduling policies to control not only age but a more sophisticated end-to-end distortion criterion by using age as an auxiliary parameter \cite{sun2019sampling, ornee2021sampling,  bedewy2020optimizing, shisher2023learning, shisher2022does}. While strikingly more demanding of analysis, these formulations take us closer to goal-oriented communication system design. \par
Almost all previous studies on the ``generate-at-will'' model adopted an assumption that the penalty of information aging is a monotonic non-decreasing function of the AoI \cite{sun2019sampling, ornee2021sampling, sun2017update, kadota2018scheduling, kadota2019scheduling, sun2019sampling2, tripathi2019whittle, klugel2019aoi, sun2019closed, bedewy2021optimal, bedewy2020optimizing}. However, it was shown in \cite{shisher2021age, shisher2022does, shisher2023learning} that the monotonicity of information aging depends heavily on the divergence of the time-series data from being a Markov chain. If the input and target data sequences in a system can be closely approximated as a Markov chain, then the penalty increases as the AoI grows; otherwise, if the data sequence significantly deviates from a Markovian structure, then the monotonicity assumption does not hold. The most closely related work to this paper is \cite{shisher2022does}, which developed scheduling policies for remote inference considering the possibly non-monotonic dependency of the AoI and practical performance. The paper \cite{shisher2022does} considered random IID delay for packet transmissions from the transmitter to the receiver and assumed a delay-free feedback channel from the receiver to the transmitter. In practical scenarios, there may be a feedback delay \cite{tsai2021unifying, pan2022optimal}, and this may affect the performance of the remote inference algorithm. Moreover, in the light of the application scenarios mentioned above (e.g., satellite and space communication), the IID delay may not be a good model, requiring a model that captures the memory in the delay process as in \cite{sun2017update}. Delayed feedback and significantly varying distribution of transmission delay are the main differences of this paper from \cite{shisher2022does}, and these differences require more technical efforts to solve the problem, as explained later in Section III. To that end, the following are the technical contributions of this paper:
\begin{itemize}
    \item We developed an index-based threshold policy to minimize the steady-state time-average inference error of an intelligent application whose input data packets are transmitted from a distant location. Given the age penalty function corresponding to a particular application, not necessarily monotonic, our policy ensures optimal practical performance, making the scheduler operate in a \textit{Goal-oriented} manner.
    \item The policy in \cite{shisher2022does} cannot achieve the optimal in the systems where channel delay statistics vary significantly with memory because of IID transmission delay and immediate feedback assumption. Hence, we expanded upon the formulation introduced in \cite{shisher2022does}, incorporating the Markovian transmission delay, with a finite number of states, and removing the assumption of immediate feedback. Although this extension is crucial for eliminating the limitations of the policy in \cite{shisher2022does}, it was not clear whether such a nicely structured policy still existed. The developed policy preserves the structure and provides a solution for practical systems with significantly varying delay statistics. Such systems will be abundant due to the envisioned coexistence of terrestrial and non-terrestrial connections in 6G, as mentioned above.
    \item The simulation results highlight the significance of (i) the policy being adaptive to non-monotonic age penalty functions and (ii) considering the memory of the delay. Numerically, the generate-at-will + zero-wait policy achieves $33\%$ to $57\%$ higher inference error than the developed policy. Furthermore, the performance gain from incorporating delay memory increases by up to $13\%$ with growing delay memory.
\end{itemize}

\section{System Model and Problem Formulation}
\begin{figure}[t]
\centerline{\includegraphics[width=.5\textwidth]{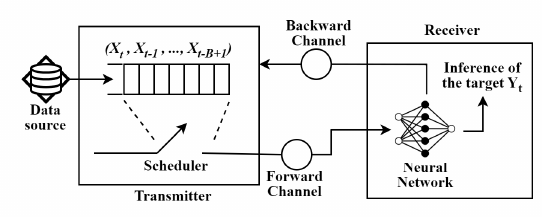}}
\caption{The remote inference system with two-way communication where the neural network on the receiver side predicts the current value of a target variable using the most recently delivered packet.}
\label{system_Model}
\end{figure}

We consider a discrete-time remote inference system, as depicted in Fig. \ref{system_Model}. The transmitter incorporates a data buffer and a scheduler. At each time slot $t$, a new data packet $X_{t} \in \mathcal{X}$ is sampled from the source and added to the buffer; meanwhile, the oldest is discarded. By this, the buffer contains the most recently sampled $B$ packets $(X_{t}, X_{t-1}, ..., X_{t-B+1})$ at any time slot $t$. The scheduler can submit any packet stored in the buffer to the channel at suitable time slots. The Age of Information (AoI), $\Delta(t)$, is the time difference between the current time $t$ and the generation time $t-\Delta(t)$ of the most recently delivered packet $X_{t-\Delta(t)}$. The value of AoI depends on (i) which packet in the buffer is selected by the scheduler during the submission of the most recently delivered packet and (ii) communication delays. A trained neural network on the receiver side takes the packet $X_{t-\Delta(t)}$ and the AoI $\Delta(t)$ as inputs and produces an action $a \in \mathcal A$ to infer the current value of a target signal $Y_{t} \in \mathcal{Y}$ at each time slot $t$. Given the AoI $\Delta(t)=\delta$, the average inference error is \cite{shisher2022does}
\begin{equation} \label{eqn:0}
    h(\delta) = \mathbb{E}_{Y, X  \sim P_{Y_{t}, X_{t-\delta}}}[L(Y,\phi(\delta, X_{t-\delta})],
\end{equation}
where $\phi:\mathbb{Z^{+}}\times\mathcal{X}\mapsto\mathcal{A}$ is the function representing the neural network, $L:\mathcal{Y}\times\mathcal{A}\mapsto\mathbb{R}$ is a loss function, and $P_{Y_{t}, X_{t-\delta}}$ is the joint distribution of the target $Y_{t}$ and the packet $X_{t-\delta}$. $L(y,a)$ is the incurred loss if the output $a$ is used for prediction when the target signal $Y_{t}=y$.\par
The loss function $L$ is determined by the purpose of the application. For example, a quadratic loss function $L_{2}(y,\hat{y})=\lVert y-\hat{y} \rVert_{2}^{2}$ is used in neural network based minimum mean-squared estimation, where the action $a=\hat{y}$ is an estimate of the target $Y_{t}=y$ and $\lVert y \rVert_{2}$ is the Euclidean norm. In softmax regression (i.e., neural network based maximum likelihood classification), the action $a=\mathcal{Q}_{Y}$ is a distribution of $Y_{t}$, and the loss function $L_{log}(y,\mathcal{Q}_{Y})=-log\mathcal{Q}_{Y}(y)$ is the negative log-likelihood function of the value $Y_{t}=y$. \par
The communication is carried out between the transmitter and receiver over a two-way channel. The two-way channel has a finite number of states $\mathcal{C} = (1, 2, ..., C)$. Each state $c$ corresponds to a distinct transmission and feedback delay distribution represented by the discrete random variables $Q_{c}$ and $R_{c}$, respectively, where $\mathbb{E}[Q_{c}]<\infty \text{ }$ and $\mathbb{E}[R_{c}]<\infty$ for all $c$. The delay duration is at least one time slot for each packet or feedback transmission. In practice, the state $c$ of the channel could be used to characterize time-varying physical phenomena that affect the transmission and feedback delay distribution, $Q_c$ and $R_c$. Examples of such phenomena include (i) the transmission duration varying due to an erasure channel situation leading to possibly a number of retransmissions, (ii) long or short network routes that packets may go through, and (iii) the delay increasing or decreasing with network congestion varying in time. Upon the delivery of the $i$-th packet, the receiver is able to detect the channel state $c_{i}$ valid during the transmission of packet $i$. Then, the channel state $c_{i}$ is fed back to the transmitter along with an ACK. The two-way channel is reliable; that is, no packet or message is lost during the transmission.\par
Recent studies \cite{shisher2022does, ShisherTimely} have exhibited scenarios where the inference error in \eqref{eqn:0} is not monotonic with age, where fresher packets do not always lead to a smaller inference error than stale ones. In other words, in certain scenarios, an older packet with AoI $>0$ can be more relevant than a fresher packet, in the sense that this packet can lead to a smaller inference error than a packet with {AoI~ $= 0$} \cite{shisher2022does, ShisherTimely}. In these cases, the "generate-at-will" approach, which samples a fresh packet upon making a transmission decision, will not be general enough to achieve optimal performance. In recognition of this fact, the "selection-from-buffer" model was proposed in \cite{shisher2022does, shisher2023learning}. In this paper, we adopt this latter model. \par
The system starts to operate at time slot $t=0$. We assume that the buffer initially contains the packets $(X_{0}, X_{-1}, ..., X_{-B+1})$. By this, the buffer is kept full at all time slots $t\geq 0$. At time slot $S_i$, the packet $X_{S_i-b_i}$ is submitted to the channel, which is the $(b_i + 1)$-th freshest packet in the buffer, with $b_i \in \mathcal{B}=\{0, 1, \ldots, B-1\}$. The packet is delivered to the receiver at time slot $D_{i}$, and the transmitter receives the corresponding ACK and channel state $c_i$ at time slot $A_{i}$, such that $S_i < D_i < A_i$. We assume that the transmitter always waits for the ACK before submitting a new packet to the channel. In other words, the scheduler always remains silent at the time slots between $S_{i}$ and $A_{i}$ for all $i$. Let $i$-th epoch be composed of the time slots in the interval $[A_{i-1}, A_{i})$. The channel state $c_{i}$ specifies the transmission delay distribution in the $i$-th epoch and evolves following a finite-state ergodic Markov chain with the transition probabilities $p_{ij}$, where $i,j\in \mathcal{C}$, and $p_{ij}$ is the transition probability from state $i$ to state $j$. The Markov chain makes a single transition at the time slots $A_{i}$ and none otherwise. Let $T_{i}=D_{i}-S_{i}$ and $F_{i}=A_{i}-D_{i}$ be the incurred transmission and feedback delays in $i$-th epoch, respectively. The distributions of $T_{i}$ and $F_{i}$ are determined by the channel state in the $i$-th epoch. That is, $p_{T_i|(c_i=c)}(q)=p_{Q_c}(q)$ and $p_{F_i|(c_i=c)}(r)=p_{R_c}(r)$, where $p_{K}(k)$ is the probability mass function (PMF) of the discrete random variable $K$. The AoI at time slot $t$ is given by
\begin{equation} \label{eqn:1}
    \Delta(t) = t-S_{i}+b_{i}, \text{  if  } D_{i} \leq t < D_{i+1}. 
\end{equation}
The initial conditions of the system are $S_{0}=0$, $\Delta(0)<\infty$, $c_{0} \in \mathcal{C}$, and $b_0 \in \mathcal{B}$. \par
We strive to minimize the expected inference error per time slot in steady-state. Accordingly, the scheduler on the transmitter side has to determine (i) when to submit a packet to the channel and (ii) which packet in the buffer to submit. A scheduling policy is a tuple $(f, g)$, where the buffer position sequence $f=(b_{1}, b_{2}, ...)$ controls which packet in the buffer to submit in each epoch, and the packet submission time sequence $g=(S_{1}, S_{2}, ...)$ specifies when to submit the packet to the channel and start the transmission in each epoch. Let $\Pi$ be the set of all policy tuples $(f, g)$. Then, the problem formulation is given by
\begin{equation} \label{eqn:2}
    h_{opt}=\inf_{(f,g)\in \Pi}\limsup_{T\to\infty} \frac{1}{T}\mathbb{E}_{(f,g)}\Bigg[\sum_{t=0}^{T-1} h(\Delta(t))\Bigg],
\end{equation}
where $h(\Delta(t))$ is the inference error at time slot $t$, and $h_{opt}$ is the optimum value of \eqref{eqn:2}. \par
This problem formulation extends the single-source scheduling problem in \cite{shisher2022does} to more practical systems by considering the memory of the transmission delay and by removing the delay-free feedback assumption.

\section{Optimal Scheduling Policy}
We provide an optimal scheduling policy to (\ref{eqn:2}) in two steps. First, we consider a class of policy tuples $(f_{\psi},g) \in \Pi$ such that $f_\psi = (b_1,b_2,\ldots)$ is a stationary deterministic buffer position sequence, where
\begin{equation} \label{eqn:3}
    b_{i+1} = \psi(c_{i}),
\end{equation} for all $i = \{0,1,2,\ldots\}$, and $\Psi$ is the set of all mapping functions $\psi: \mathcal C \mapsto \mathcal B$. Given a stationary deterministic buffer position sequence $f_\psi = (\psi(c_0), \psi(c_1), \ldots ) $, we first optimize the packet submission time sequence $g = (S_1,S_2,\ldots)$: 
\begin{equation} \label{eqn:4}
    h_{\psi,opt}=\inf_{(f_{\psi},g)\in \Pi}\limsup_{T\to\infty} \frac{1}{T}\mathbb{E}_{(f_{\psi},g)}\Bigg[\sum_{t=0}^{T-1} h(\Delta(t))\Bigg],
\end{equation}
where $h_{\psi,opt}$ is the optimum value of \eqref{eqn:4}. In the second step, we present an optimal scheduling policy for the original problem \eqref{eqn:2} using the solution to \eqref{eqn:4}. \par
The optimal packet submission time sequence $g$ for \eqref{eqn:4} will be described by using an index function defined by
\begin{equation} \label{eqn:8}
    \gamma(\delta, c)=\inf_{\nu\in\mathbb{Z}^{+}} \frac{1}{\nu}\sum_{k=0}^{\nu-1}\mathbb{E}\bigg[h(\delta+T_{i+1}+k)\bigg|c_{i}=c\bigg],
\end{equation}
where $\delta \in \mathbb{Z}^{+}$ and $c \in \mathcal{C}$.
\begin{theorem} \label{th:1}
    If $|h(\delta)|<M$ for all $\delta = 1,2,\ldots$, then the  packet submission time sequence $g=(S_{1}(\beta_{\psi}), S_{2}(\beta_{\psi}), ...)$ is an optimal solution to \eqref{eqn:4}, where
    \begin{equation} \label{eqn:10}
        S_{i+1}(\beta_{\psi}) = \min\{t\geq A_{i}(\beta_{\psi}):\gamma(\Delta(t), c_{i}) \geq \beta_{\psi}\},
    \end{equation} and $\beta_{\psi}$ is the unique root of
    \begin{equation} \label{eqn:11}
        \mathbb{E}\Bigg[\sum_{t=A_{i}(\beta_{\psi})}^{A_{i+1}(\beta_{\psi})-1} h(\Delta(t))\Bigg]-\beta_{\psi}\mathbb{E}[A_{i+1}(\beta_{\psi})-A_{i}(\beta_{\psi})] =0,
    \end{equation}
    where $A_{i}(\beta_{\psi})=S_{i}(\beta_{\psi})+T_i+F_i$ is the $i$-th ACK feedback time and $\Delta(t)=t-S_{i}(\beta_{\psi})+b_i$ is the AoI at time slot $t$. Moreover, $\beta_{\psi}$ is exactly the optimum value of (\ref{eqn:4}), i.e., $\beta_{\psi}=h_{\psi,opt}$.
\end{theorem}
\begin{proof}[Proof sketch]
The problem \eqref{eqn:4} is an infinite-horizon average-cost semi-Markov decision process (SMDP) \cite[Chapter 5.6]{bertsekasdynamic}.  Define $\tau=S_{i+1}-A_i$ as the waiting time for sending the $(i+1)$-th packet after the ACK feedback of the $i$-th packet is delivered to the transmitter. Given $\Delta(A_i)=\delta$ and $c_i=c$, the Bellman optimality equation of the  average-cost SMDP \eqref{eqn:4} is 
\begin{align}\label{relativeValuederived}
&V_{\psi}(\delta, c)=\nonumber\\
&\inf_{\tau \in \{0, 1, \ldots\}}~\mathbb E \left [ \sum_{k=0}^{\tau+T_{i+1}-1} \bigg(h(\delta+k) -h_{\psi,opt}\bigg) \bigg| c_i=c\right]\nonumber\\
&+\mathbb E \left [ \sum_{j=0}^{F_{i+1}-1} \bigg(h(\psi(c_i)+T_{i+1}+j) -h_{\psi,opt}\bigg) \bigg| c_i=c\right]\nonumber\\
&+\mathbb E[V_{\psi}(\psi(c_i)+T_{i+1}+F_{i+1}, c_{i+1})| c_i=c], 
\end{align}
for all $\delta \in \mathbb{Z}^{+}$, $\psi(\cdot) \in \Psi$, and $c \in \mathcal C$, where $V_{\psi}(\cdot)$ is the relative value function of the average-cost SMDP \eqref{eqn:4}.
Theorem \ref{th:1} is proven by directly solving the Bellman optimality equation \eqref{relativeValuederived}. The details are provided in Appendix A.
\end{proof}
Theorem \ref{th:1} signifies that the optimal solution to \eqref{eqn:4} is an index-based threshold policy, where the index function depends on both the AoI and the channel state in the previous epoch. The scheduler submits $(i+1)$-th packet according to \eqref{eqn:3} at the first time slot for which the following conditions are satisfied: (i) the ACK of the previous transmission is received, i.e., $t\geq A_{i}(\beta_{\psi})$, and (ii) the index function exceeds the threshold. The threshold $\beta_{\psi}$ is exactly the optimum value of \eqref{eqn:4} and can be found by solving equation \eqref{eqn:11} using the low-complexity algorithms, e.g., bisection search, provided in \cite[Algorithms~1-3]{ornee2021sampling}. The index function in \eqref{eqn:8} is obtained by directly solving the Bellman optimality equation \eqref{relativeValuederived}. \par
Recent studies \cite{shisher2022does, shisher2023learning} have revealed a connection between the AoI optimization problems, such as \eqref{eqn:4}, and the \emph{restart-in-state} formulation of the Gittins index in \cite[Ch.~2.6.4]{gittins2011multi}. The optimal scheduling policy in Theorem \ref{th:1} has been obtained by employing that connection. \par
Now, we present an optimal solution to problem \eqref{eqn:2}.
\begin{theorem} \label{Th:2}
    If $|h(\delta)|<M$ for all $\delta = 1,2,\ldots$, then the policy tuple $(f^*,g^*)$ is an optimal solution to \eqref{eqn:2} such that
    \begin{itemize}
        \item[i.)] $f^*=(b^*_{1}, b^*_{2}, \ldots)$, where
        \begin{align}
            b^*_{i+1} &= \psi^*(c_{i}) \text{, } i=0,1,2,\ldots, \\
            \psi^* &= arg\min_{\psi \in \Psi} h_{\psi,opt},
        \end{align} and $h_{\psi,opt}$ is the optimum value of \eqref{eqn:4}.
        \item[ii.)] $g^*=(S^*_{1}, S^*_{2}, \ldots)$, where
        \begin{multline} \label{eqn:29}
            S^*_{i+1} = \min\{t\geq S^*_{i}+T_{i}+F_{i}:\gamma(\Delta(t), c_{i}) \geq h_{opt}\},
        \end{multline} and
        \begin{equation} \label{eqn:30}
            h_{opt}=\min_{\psi \in \Psi} h_{\psi,opt}.
        \end{equation}
    \end{itemize}
\end{theorem}
\begin{proof} 
See Appendix B.
\end{proof} \par
Theorem \ref{Th:2} points out that the optimal buffer position at any submission time slot is independent of the current AoI and only depends on the channel state in the previous epoch. The reasoning behind such a result can be understood deeply by analyzing the system. At the beginning of epoch $[D_{i}, D_{i+1})$, i.e., at time slot $D_{i}$, the AoI of the process resets to $b_{i}+T_{i}$, indicating that the buffer position $b_{i}$ determines the AoI reset value for time slot $D_{i}$. The epoch $[D_{i}, D_{i+1})$ evolves starting from this AoI value. Then, the AoI resets again at time slot $D_{i+1}$ following the delivery of the subsequent packet, and the AoI values in epoch $[D_{i}, D_{i+1})$ become irrelevant to the future evolution of the process. In other words, each epoch of the process has an isolated AoI evolution, and $b_{i}$ is the parameter that allows us to control AoI evolution in $i$-th epoch. Apart from $b_{i}$, $T_{i}$ is the other parameter affecting the AoI reset value for time slot $D_{i}$, and $c_{i-1}$ reveals the distribution of $T_{i}$ thanks to the Markovian structure. Thus, $b_{i}$ is chosen, considering the distribution of $T_{i}$, through $c_{i-1}$, to ensure the AoI of the process reset to the desired value at time slot $D_{i}$. Due to this inherent structure of the process, as shown in Theorem \ref{Th:2}, the optimal buffer position sequence to problem \eqref{eqn:2} is $f^*=f_{\psi^*}=(\psi^*(c_0), \psi^*(c_1), \ldots )$, where $\psi^* \in \Psi$ is the optimal mapping function. Since $\psi^* \in \Psi$, the optimal packet submission time sequence is obtained by Theorem \ref{th:1}. \par
Even though the optimal policies outlined in Theorems \ref{th:1} and \ref{Th:2} maintain the nice structure of the policies in \cite{shisher2022does}, there are two critical distinctions. Firstly, the derived index function for this problem formulation is not solely dependent on the AoI but also incorporates the channel state. The existence of such an index function was not evident in the presence of a finite number of channel states, necessitating additional technical efforts for derivation. Secondly, in contrast to \cite{shisher2022does}, where the optimal buffer position sequence is constant due to the assumption of IID transmission delay and immediate feedback, this problem formulation considers transmission delay and non-zero feedback delay significantly varying with memory. Consequently, achieving optimal performance requires a buffer position sequence that can adapt to these variations in delay statistics. This paper unveils the relationship between the buffer position sequence and delay memory while developing Theorems \ref{th:1} and \ref{Th:2}. \par
In the special case that the inference error $h(\delta)$ is a non-decreasing function of the AoI $\delta$, the index function $\gamma(\delta,c)$ in \eqref{eqn:8} reduces to $\mathbb{E}[h(\delta+T_{i+1})|c_{i}=c]$. Moreover, the optimal buffer position sequence is $f^*=(0,0,\ldots)$.

\begin{figure}[t]
\centerline{\includegraphics[width=.5\textwidth]{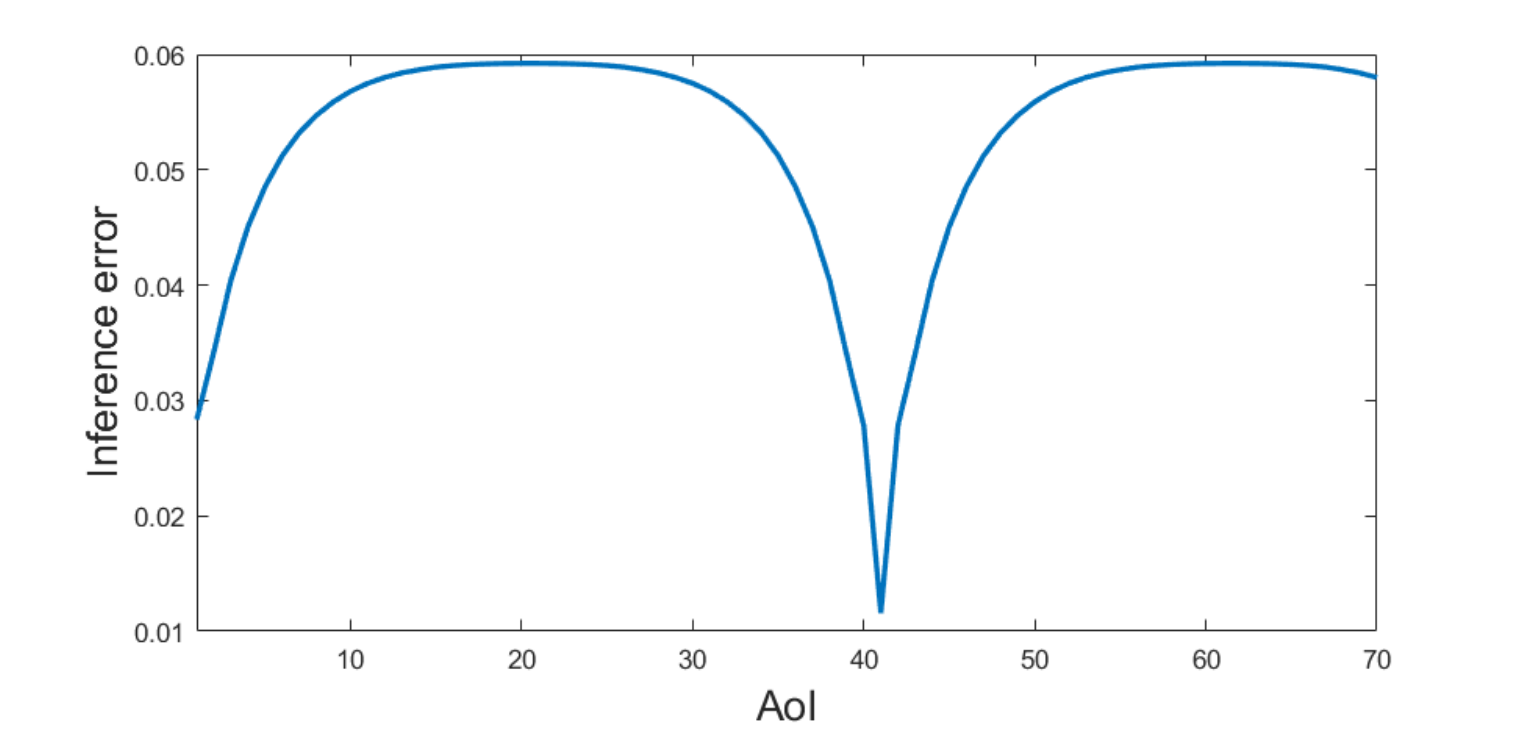}}
\caption{The incurred inference error when the current sample of the target $Y_{t}$ is predicted using a past sample $X_{t-\delta}$ with AoI $\delta$ ranging from 1 to 70.}
\label{IEvsAoI}
\end{figure}

\section{Simulation Results}
In this section, we demonstrate the performance of our optimal scheduling policy in Theorem \ref{Th:2}. To obtain an inference error function $h(\delta)$, we use a discrete-time autoregressive model of the order $p$ (AR($p$)):
\begin{align}
X_{t}=a_1 X_{t-1}+ a_2 X_{t-2}+\ldots+a_p X_{t-p}+W_t,
\end{align}
where the noise $W_t \in \mathbb R$ is zero-mean Gaussian with variance $\sigma_W^2$ and $X_t \in \mathbb R$. Let $Y_t=X_t+N_t$ be the target variable, where the noise $N_t$ is zero-mean Gaussian with variance $\sigma_N^2$. Fig. \ref{IEvsAoI} indicates the average inference error values incurred when the target $Y_t$ is predicted using $X_{t-\delta}$, where the value of AoI $\delta$ ranges from 1 to 70. We set $\sigma_W^2=0.01$, $\sigma_N^2=0.001$ and construct an AR($50$) with coefficients $a_{38}=a_{44}=0.007$, $a_{39}=a_{43}=0.05$, $a_{40}=a_{42}=0.1$, and $a_{41}=0.68$. The remaining coefficients are zero. We consider a quadratic loss function $L_{2}(y,\hat{y})=\lVert y-\hat{y} \rVert_{2}^{2}$. Because $X_{t-\delta}$ and $Y_t$ are jointly Gaussian, and the loss function is quadratic, the optimal inference error performance is achieved by a linear MMSE estimator. Hence, a linear regression algorithm is adopted in our simulations. However, our study can be readily applied to other neural network-based predictors.\par
We compare the performance of three scheduling policies: (i) the optimal policy in Theorem \ref{Th:2}, (ii) the single-source scheduling policy in \cite{shisher2022does}, and (iii) generate-at-will + zero wait policy: $(f,g)$ is such that $f=(0, 0, \ldots)$ and $g=(A_{1}, A_{2}, \ldots)$. \par
The simulation parameters related to the channel statistics have been set as follows: (i) the number of channel states $C=2$, (ii) the random variable $Q_{1}$ has the PMF $p_{Q_{1}}(3)=0.7$, and $p_{Q_{1}}(4)=0.3$, (iii) the random variable $Q_{2}$ has the PMF $p_{Q_{2}}(11)=0.35$, and $p_{Q_{2}}(12)=0.65$, (iv) the random variable $R_{1}$ has the PMF $p_{R_{1}}(1)=1$, and (v) the random variable $R_{2}$ has the PMF $p_{R_{2}}(4)=1$. The memory of the transmission delay is adjusted by varying the parameter $\alpha=p_{12}+p_{21}$ between $0$ and $2$. We set $p_{12}=p_{21}$, which ensures that the fraction of epochs with convenient delay condition is always $0.5$. The scheduling policy in \cite{shisher2022does} assumes both channel states are equally likely at each epoch independent of the history for any value of $\alpha$.\par
\begin{figure}[t]
\centerline{\includegraphics[width=.5\textwidth]{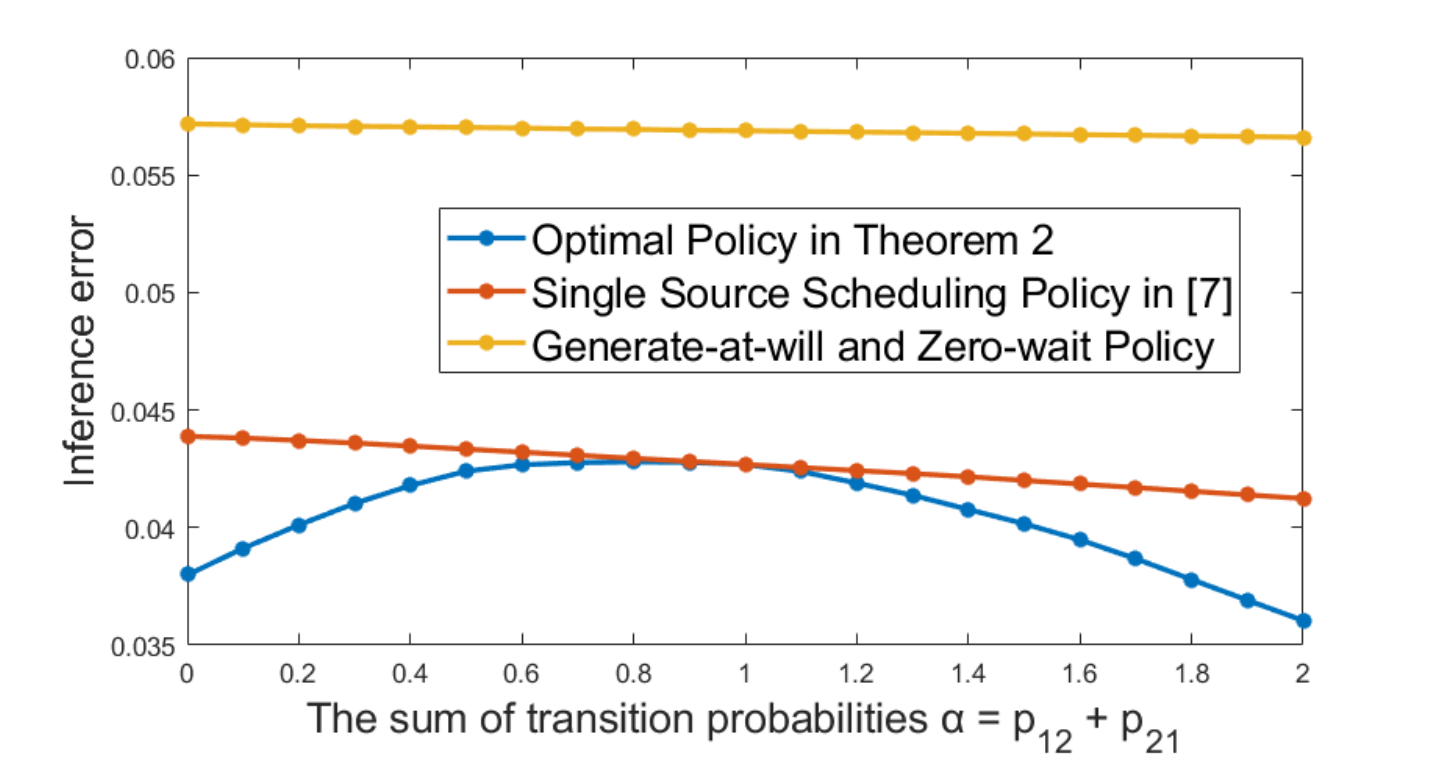}}
\caption{The time-average inference error achieved by the three different scheduling policies.}
\label{final_Result}
\end{figure}

Fig. \ref{final_Result} presents the time-average inference error values achieved by the three scheduling policies mentioned above. The results underscore that consistently submitting the freshest packet with zero waiting time leads to a $33\%$ to $57\%$ higher inference error compared to the optimal policy outlined in Theorem \ref{Th:2}. Additionally, the optimal policy shows minimal improvement compared to the corresponding policy in \cite{shisher2022does} for $\alpha$ values around $1$. The reason for the negligible improvement is that the delay distribution is exactly IID when $\alpha=1$. However, as $\alpha$ deviates from this point, resulting in increased delay memory, the advantage of the optimal policy described in Theorem \ref{Th:2} becomes apparent, leading to an important performance gain of up to $13\%$.

\section{Conclusion}
In this paper, we studied a remote inference problem where a neural network on the receiver side predicts the real-time value of a target signal using the data packets transmitted from a distant location. Motivated by the foreseen coexistence of terrestrial and non-terrestrial connections in 6G, we assumed it would be typical for a data flow to be served through multiple alternative paths and considered transmission and feedback delay distributions significantly varying with memory. For this system model, we developed an optimal index-based threshold policy that minimizes the expected inference error per time slot in steady-state. The policy optimizes the performance for a given inference error function specific to the application and goal on the receiver side, addressing the expectation for future networks to operate in a goal-oriented manner. Finally, we demonstrated, with the simulation results, the performance gain that can be achieved in the remote inference problems by considering the memory of the delay.

\bibliographystyle{IEEEtran}
\bibliography{ISIT}

\section*{Appendix A: Proof of Theorem 1}
The scheduling problem \eqref{eqn:4} is an infinite-horizon average-cost semi-Markov decision process (SMDP) \cite[Chapter 5.6]{bertsekasdynamic}. According to \cite[Chapter 5.6]{bertsekasdynamic}, we describe the components of this problem in detail below:

\begin{itemize}
    \item{\bf Decision Time:} Each $i$-th ACK feedback time $A_i=S_i+T_i+F_i$ is a decision time of the problem \eqref{eqn:4}, where $S_i$ is the submission time of the $i$-th packet, which takes $T_i\geq 1$ time slots to be delivered to the receiver, and the $i$-th ACK feedback takes $F_i\geq 1$ time slots to be delivered to the transmitter.
     \item {\bf State:} At time $A_i$, the state of the system is represented by $(\Delta(A_i), c_i)$, where $\Delta(A_i)$ is the AoI at time slot $A_i$ and $c_i$ is the channel state in $i$-th epoch. 
     \item {\bf Action:} Let $\tau_i=S_{i+1}-A_i$ represent the waiting time for sending the $(i+1)$-th packet after the $i$-th ACK feedback is received. As we consider $S_0=0$, $S_i=\sum_{j=0}^{i-1} (T_j+F_j+\tau_j)$ and $A_i=S_i+T_i+F_i$ for each $i \in \{0,1,\ldots\}$. Given $(T_0, T_1, \ldots)$ and $(F_0, F_1, \ldots)$, the sequence $(S_1, S_2, \ldots)$ is uniquely determined by $(\tau_0, \tau_1, \ldots)$. Hence, one can also use $(\tau_0, \tau_1, \ldots)$ to represent a sequence of actions instead of $(S_1, S_2, \ldots)$. The scheduler decides the waiting time $\tau_i$ at time slot $A_i$. Then, the packet in the buffer position $b_{i+1}=\psi(c_i)$ is submitted to the channel at time slot $S_{i+1}$.
     \item {\bf State Transitions:} The AoI $\Delta(t)$ evolves as follows:
    \begin{align}\label{AoIProcess}
    \!\! \Delta(t)=
      \begin{cases}
        T_{i}+b_i, &\text{if}~t=D_{i} \\
        \Delta(t-1)+1, &\text{otherwise}
    \end{cases}
    \end{align} for $i \in \{0,1, \ldots\}$.
    The state of the channel in $i$-th epoch, $c_{i}$, evolves following a finite-state ergodic Markov chain with the transition probabilities $p_{ij}$, where $i,j\in \mathcal{C}$, and $p_{ij}$ is the transition probability from state $i$ to state $j$. The Markov chain makes a single transition at the time slots $A_{i}$ and none otherwise.
    \item {\bf Expected Transition Time:} The expected time difference between the decision times $A_i$ and $A_{i+1}$ given $c_i=c$ is $\mathbb E[A_{i+1}-A_i|c_i=c] =\mathbb E[\tau_i+T_{i+1}+F_{i+1}|c_i=c].$
    \item {\bf Expected Transition Cost:} Given $c_i=c$, the expected cumulative cost incurred during the interval between the decision times $A_i$ and $A_{i+1}$ can be calculated as
    \begin{align}\label{interdeliverycost}
    &\mathbb E\left[\sum_{t=A_i}^{A_{i+1}-1}h(\Delta(t))\bigg|c_i=c\right]=\nonumber\\
    &\mathbb E\left[\sum_{k=0}^{\tau_{i}+T_{i+1}-1} h(\Delta(A_i)+k)\bigg|c_i=c\right]\nonumber\\
    &+\mathbb E\left[\sum_{k=0}^{F_{i+1}-1} h(\psi(c_i)+T_{i+1}+k)\bigg|c_i=c\right]
    \end{align}
\end{itemize}
The average-cost SMDP \eqref{eqn:4} can be solved by using dynamic programming \cite{bertsekasdynamic, puterman2014markov}. Given $\Delta(A_i)=\delta$ and $c_i=c$, the Bellman optimality equation of the average cost problem is  
\begin{align}\label{Bellman2}
&V_{\psi}(\delta, c)= \nonumber\\
&\inf_{\tau_{i} \in \{0, 1, \ldots\}}\mathbb E \left [ \sum_{k=0}^{\tau_{i}+T_{i+1}-1} \bigg(h(\delta+k) -h_{\psi,opt}\bigg) \bigg| c_i=c\right]\nonumber\\
&+\mathbb E \left [ \sum_{k=0}^{F_{i+1}-1} \bigg(h(\psi(c_i)+T_{i+1}+k) -h_{\psi,opt}\bigg) \bigg| c_i=c\right]\nonumber\\
&+\mathbb E[V_{\psi}(\psi(c_i)+T_{i+1}+F_{i+1}, c_{i+1})| c_i=c],
\end{align}
where $V_{\psi}(\delta, c)$ is the relative value function associated with the average cost problem \eqref{eqn:4}. \par
Let $\bar{\tau}=\tau(\delta, c, h_{\psi,opt})$ be an optimal solution to \eqref{Bellman2}. If $\Delta(A_i)=\delta$ and $c_i=c$, then the optimal waiting time $\tau_i$ for sending the $(i+1)$-th packet is $\bar{\tau}$. 
Because the terms $$\mathbb E \left [ \sum_{k=0}^{F_{i+1}-1} \bigg(h(\psi(c_i)+T_{i+1}+k) -h_{\psi,opt}\bigg) \bigg| c_i=c\right]$$ and $$\mathbb E[V_{\psi}(\psi(c_i)+T_{i+1}+F_{i+1}, c_{i+1})| c_i=c]$$
do not depend on the waiting time $\bar{\tau}$, the optimization problem can be reformulated as 
\begin{align}\label{Bellman3}
\inf_{\bar{\tau} \in \{0, 1, \ldots\}}\mathbb E \left [ \sum_{k=0}^{\bar{\tau}+T_{i+1}-1} \bigg(h(\delta+k) -h_{\psi,opt}\bigg) \bigg| c_i=c\right].
\end{align}
From \eqref{Bellman3}, we can show that $\bar{\tau}=0$ if 
\begin{align}\label{inequality1}
&\inf_{\bar{\tau} \in \{1, 2, \ldots\}}\mathbb E \left [ \sum_{k=0}^{\bar{\tau}+T_{i+1}-1} \bigg(h(\delta+k) -h_{\psi,opt}\bigg) \bigg| c_i=c\right] \nonumber\\
&\geq \mathbb E \left [ \sum_{k=0}^{T_{i+1}-1} \bigg(h(\delta+k) -h_{\psi,opt}\bigg) \bigg| c_i=c\right].
\end{align}
After some rearrangement, the inequality \eqref{inequality1} can also be expressed as 
\begin{align}\label{inequality2}
\inf_{\bar{\tau} \in \{1, 2, \ldots\}}\mathbb E \left [ \sum_{k=0}^{\bar{\tau}-1} \bigg(h(\delta+T_{i+1}+k) -h_{\psi,opt}\bigg) \bigg| c_i=c\right] \geq 0.
\end{align}
The inequality \eqref{inequality2} holds if and only if
\begin{align}\label{inequality3}
\inf_{\bar{\tau} \in \{1, 2, \ldots\}} \frac{1}{\bar{\tau}} \mathbb{E} \left[ \sum_{k=0}^{\bar{\tau}-1} h(\delta+T_{i+1}+k)\bigg| c_i=c \right] \geq h_{\psi,opt}.
\end{align}
The left-hand side of \eqref{inequality3} is equal to the index function $\gamma(\delta, c)$ given by \eqref{eqn:8}. Similarly, $\bar{\tau}=1$, if $\bar{\tau}\neq 0$ and 
\begin{align}\label{inequality4}
\inf_{\bar{\tau} \in \{2, 3, \ldots\}}\mathbb E \left [ \sum_{k=1}^{\bar{\tau}-1} \bigg(h(\delta+T_{i+1}+k) -h_{\psi,opt}\bigg) \bigg| c_{i}=c\right] \geq 0.
\end{align}
The inequality \eqref{inequality4} can be rewritten as 
\begin{align}\label{step2}
\gamma(\delta+1, c)\geq h_{\psi,opt}.
\end{align}
By repeating the same steps as \eqref{inequality4}-\eqref{step2}, we can obtain $\bar{\tau}=k$ is optimal, if $\bar{\tau}\neq 0, 1, \ldots, k-1$ and 
\begin{align}
\gamma(\delta+k, c)\geq h_{\psi,opt}.
\end{align}
Hence, the optimal waiting time $\tau_i=\bar{\tau}$ is determined by
\begin{align}\label{optimalwaitingb}
\bar{\tau}=\tau(\delta, c, h_{\psi,opt})=\min\{k \geq 0: \gamma(\delta+k, c)\geq h_{\psi,opt}\}.
\end{align} \par
Next, we need to compute the optimal objective value $h_{\psi,opt}$. Let $\varphi_{i}$ denote the random variable $\psi(c_{i-1})+T_{i}$. Because $\Delta(A_i(\beta_{\psi}))=\varphi_{i}+F_{i}$ and $A_{i+1}(\beta_{\psi})=A_{i}(\beta_{\psi})+\tau(\varphi_{i}+~F_{i}, c_i,\beta_{\psi})+T_{i+1}+F_{i+1}$, we can write
\begin{align}\label{proofeq1}
&\mathbb{E}\Bigg[\sum_{t=A_{i}(\beta_{\psi})}^{A_{i+1}(\beta_{\psi})-1} h(\Delta(t))\Bigg]-\beta_{\psi}\mathbb{E}[A_{i+1}(\beta_{\psi})-A_{i}(\beta_{\psi})]= \nonumber\\
&\mathbb{E}\Bigg[\sum_{k=0}^{\tau(\varphi_{i}+F_{i}, c_i,\beta_{\psi})+T_{i+1}-1} h(\varphi_{i}+F_{i}+k)\Bigg] \nonumber\\
&+\mathbb{E}\Bigg[\sum_{k=0}^{F_{i+1}-1} h(\varphi_{i+1}+k)\Bigg]\nonumber\\
&-\beta_{\psi}\mathbb{E}[\tau(\varphi_{i}+F_{i}, c_i,\beta_{\psi})+T_{i+1}+F_{i+1}].
\end{align}
Equation \eqref{proofeq1} implies that $h_{\psi,opt}=\beta_{\psi}$ is the root of \eqref{eqn:11}, if the following holds:
\begin{align}\label{proofeq2}
&\mathbb{E}\Bigg[\sum_{k=0}^{\tau(\varphi_{i}+F_{i}, c_i,h_{\psi,opt})+T_{i+1}-1} h(\varphi_{i}+F_{i}+k)\Bigg]\nonumber\\
&+\mathbb{E}\Bigg[\sum_{k=0}^{F_{i+1}-1} h(\varphi_{i+1}+k)\Bigg]\nonumber\\
&-h_{\psi,opt}\mathbb{E}[\tau(\varphi_{i}+F_{i}, c_i,h_{\psi,opt})+T_{i+1}+F_{i+1}]=0.
\end{align}
From \eqref{Bellman2}, we get
\begin{align}\label{proofeq3}
&V_{\psi}(\varphi_{i}+F_{i}, c_i)= \nonumber\\
&\mathbb E \left [ \sum_{k=0}^{\tau(\varphi_{i}+F_{i}, c_i,h_{\psi,opt})+T_{i+1}-1} \bigg(h(\varphi_{i}+F_{i}+k) -h_{\psi,opt}\bigg) \bigg| c_i\right]\nonumber\\
&+\mathbb E \left [ \sum_{k=0}^{F_{i+1}-1} \bigg(h(\varphi_{i+1}+k) -h_{\psi,opt}\bigg) \bigg| c_i\right]\nonumber\\
&+\mathbb E[V_{\psi}(\varphi_{i+1}+F_{i+1}, c_{i+1})|c_i].
\end{align}
By the law of iterated expectations, if we take expectation over $c_i$ on both sides of \eqref{proofeq3}, we obtain
\begin{align}\label{proofeq4}
&\mathbb E[V_{\psi}(\varphi_{i}+F_{i}, c_i)]= \nonumber\\
&\mathbb E \left [ \sum_{k=0}^{\tau(\varphi_{i}+F_{i}, c_i,h_{\psi,opt})+T_{i+1}-1} \bigg(h(\varphi_{i}+F_{i}+k) -h_{\psi,opt}\bigg)\right]\nonumber\\
&+\mathbb E \left [ \sum_{k=0}^{F_{i+1}-1} \bigg(h(\varphi_{i+1}+k) -h_{\psi,opt}\bigg)\right]\nonumber\\
&+\mathbb E[V_{\psi}(\varphi_{i+1}+F_{i+1}, c_{i+1})].
\end{align}
Since $c_i$ evolves following a finite-state ergodic Markov chain with a unique stationary distribution, $\mathbb E[V_{\psi}(\varphi_{i}+F_{i},  c_i)]=\mathbb E[V_{\psi}(\varphi_{i+1}+F_{i+1}, c_{i+1})]$. Hence, we get
\begin{align}\label{proofeq5}
&\mathbb E \left [ \sum_{k=0}^{\tau(\varphi_{i}+F_{i}, c_i,h_{\psi,opt})+T_{i+1}-1} \bigg(h(\varphi_{i}+F_{i}+k) -h_{\psi,opt}\bigg)\right]\nonumber\\
&+\mathbb E \left [ \sum_{k=0}^{F_{i+1}-1} \bigg(h(\varphi_{i+1}+k) -h_{\psi,opt}\bigg)\right]=0,
\end{align}
which yields equation \eqref{proofeq2}. \par
Finally, we need to show that the root of equation \eqref{eqn:11} is unique. The right-hand side of \eqref{proofeq1} can be expressed as
\begin{align}\label{unique_root_1}
&\mathbb E \left [ \sum_{k=0}^{\tau(\varphi_{i}+F_{i}, c_i,\beta_{\psi})+T_{i+1}-1} \bigg(h(\varphi_{i}+F_{i}+k) -\beta_{\psi}\bigg)\right]\nonumber\\
&+\mathbb E \left [ \sum_{k=0}^{F_{i+1}-1} \bigg(h(\varphi_{i+1}+k) -\beta_{\psi}\bigg)\right]= \nonumber\\
&\inf_{\tau \in \{0,1, \ldots\}}\mathbb E \left [ \sum_{k=0}^{\tau+T_{i+1}-1} \bigg(h(\varphi_{i}+F_{i}+k) -\beta_{\psi}\bigg)\right]\nonumber\\
&+\mathbb E \left [ \sum_{k=0}^{F_{i+1}-1} \bigg(h(\varphi_{i+1}+k) -\beta_{\psi}\bigg)\right].
\end{align}
This expression is concave, continuous, and strictly decreasing in $\beta_{\psi}$ because the first term is the pointwise infimum of the linear decreasing functions of $\beta_{\psi}$ and the second term is a linear decreasing function of $\beta_{\psi}$. Therefore, since
$$\lim_{\beta_{\psi}\rightarrow -\infty}\mathbb E \left [ \sum_{k=0}^{\tau+T_{i+1}-1} \bigg(h(\varphi_{i}+F_{i}+k) -\beta_{\psi}\bigg)\right]=\infty$$ and $$\lim_{\beta_{\psi}\rightarrow \infty}\mathbb E \left [ \sum_{k=0}^{\tau+T_{i+1}-1} \bigg(h(\varphi_{i}+F_{i}+k) -\beta_{\psi}\bigg)\right]=-\infty$$ for any $\tau \geq 0$, and
$$\lim_{\beta_{\psi}\rightarrow -\infty}\mathbb E \left [ \sum_{k=0}^{F_{i+1}-1} \bigg(h(\varphi_{i+1}+k) -\beta_{\psi}\bigg)\right]=\infty$$ and $$\lim_{\beta_{\psi}\rightarrow \infty}\mathbb E \left [ \sum_{k=0}^{F_{i+1}-1} \bigg(h(\varphi_{i+1}+k) -\beta_{\psi}\bigg)\right]=-\infty,$$
equation \eqref{eqn:11} has a unique root. This completes the proof.

\section*{Appendix B: Proof of Theorem 2}
The scheduling problem \eqref{eqn:2} is an infinite-horizon average-cost semi-Markov decision process (SMDP) \cite[Chapter 5.6]{bertsekasdynamic}. The components \cite[Chapter 5.6]{bertsekasdynamic} of this problem are the same as the SMDP in the proof of Theorem \ref{th:1} with one exception. In this case, at any decision time $A_i$, the buffer position $b_{i+1}$ is not determined by an arbitrary mapping function $\psi(\cdot) \in \Psi$, and the scheduler can freely select the buffer position $b_{i+1}$ from the set $\mathcal{B}$. \par
The average-cost SMDP \eqref{eqn:2} can be solved by using dynamic programming \cite{bertsekasdynamic, puterman2014markov}. Given $\Delta(A_i)=\delta$ and $c_i=c$, the Bellman optimality equation of the average cost problem is  
\begin{align}\label{Bellman_proof_2}
&V(\delta, c)= \nonumber\\
&\inf_{\substack{\tau_{i} \in \{0, 1, 2, \ldots\} \\ b_{i+1} \in \{0,1,\ldots,B-1\}}}\mathbb E \left [ \sum_{k=0}^{\tau_{i}+T_{i+1}-1} \bigg(h(\delta+k) -h_{opt}\bigg) \bigg| c_i=c\right]\nonumber\\
&+\mathbb E \left [ \sum_{k=0}^{F_{i+1}-1} \bigg(h(b_{i+1}+T_{i+1}+k) -h_{opt}\bigg) \bigg| c_i=c\right]\nonumber\\
&+\mathbb E[V(b_{i+1}+T_{i+1}+F_{i+1}, c_{i+1})| c_i=c],
\end{align}
where $V(\delta, c)$ is the relative value function associated with the average cost problem \eqref{eqn:2}.
Because the term $$\mathbb E \left [ \sum_{k=0}^{\tau_{i}+T_{i+1}-1} \bigg(h(\delta+k) -h_{opt}\bigg) \bigg| c_i=c\right]$$
does not depend on the buffer position $b_{i+1}$, the optimal buffer position $b^*_{i+1}$ when $\Delta(A_i)=\delta$ and $c_i=c$ is 
\begin{align}\label{alternative_opt}
&b^*_{i+1}= arg\min_{b_{i+1} \in \{0,1,\ldots,B-1\}}\nonumber\\
&\mathbb E \left [ \sum_{k=0}^{F_{i+1}-1} \bigg(h(b_{i+1}+T_{i+1}+k) -h_{opt}\bigg) \bigg| c_i=c\right]\nonumber\\
&+\mathbb E[V(b_{i+1}+T_{i+1}+F_{i+1}, c_{i+1})| c_i=c].
\end{align}
Equation \eqref{alternative_opt} indicates that the optimal buffer position $b^*_{i+1}$ is independent of the AoI $\delta$ at the decision time slot $A_i$ and is a function of the channel state in the previous epoch $c_i=c$. We can construct a mapping $\psi^*:\mathcal{C}\mapsto \mathcal{B}$ such that
\begin{align}\label{mapping}
&\psi^*(c)= \nonumber\\
&arg\min_{b \in \mathcal{B}} \mathbb E \left [ \sum_{k=0}^{F_{i+1}-1} \bigg(h(b+T_{i+1}+k) -h_{opt}\bigg) \bigg| c_i=c\right]\nonumber\\
&+\mathbb E[V(b+T_{i+1}+F_{i+1}, c_{i+1})| c_i=c]
\end{align} for all $c \in \mathcal{C}$.
Therefore, in the optimal scenario, this problem is a special case of problem \eqref{eqn:4}, where the buffer position sequence $f_{\psi}$ is set to $f_{\psi^*}=(\psi^*(c_0), \psi^*(c_1), \ldots )$. Since $\psi^* \in \Psi$, the optimal packet submission time sequence $g^*=(S^*_{1}, S^*_{2}, \ldots)$ is obtained by Theorem \ref{th:1}. This completes the proof.

%% file: ISIT_final.tex
\maketitle
\begin{abstract}
We study the design of a goal-oriented sampling and scheduling strategy through a channel with highly variable two-way random delay, which can exhibit memory (e.g., Delay and Disruption Tolerant Networks). The objective of the communication is to optimize the performance of remote inference, where an inference algorithm (e.g., a trained neural network) on the receiver side predicts a time-varying target signal using the data samples transmitted by a sensor. Previous formulations to this problem either assumed a channel with IID transmission delay, neglecting feedback delay or considered the monotonic relation that the performance only gets worse as the input information ages. We show how, with delayed feedback, one can effectively exploit the knowledge about delay memory through an index-based threshold policy. This policy minimizes the expected time-average inference error that can be monotone or non-monotone in age. The index function is expressed in terms of the Age of Information (AoI) on the receiver side and a parameter regarding the distribution of subsequent transmission delay, both of which can readily be tracked.
\end{abstract}

\begin{IEEEkeywords}
Goal-Oriented Communications, Markovian Delay, Two-way Delay, Remote Inference, Age of Information, Delay with Memory, Disruption Tolerant Networks (DTN).
\end{IEEEkeywords}

\section{Introduction}
Orders of magnitude enhancement is desired in terms of performance, coverage, capacity, and energy efficiency in 6G with respect to existing cellular networks. Such striking objectives accompany the demand for intelligent applications that run based on remotely collected data, sometimes involving non-terrestrial links. Based on the envisioned coexistence of terrestrial and non-terrestrial connections in 6G, it will not be unusual for network connections to be serviced through multiple alternative paths with highly variable delay. The relays in the network may force the utilization of a selected path for a particular application, and this path may be changed in time to manage the variety of flows in the network. Therefore, while developing efficient policies to optimize the performance of intelligent applications, considering their adaptability to significantly varying delay conditions, specifically the time dependence in the delay statistics, is crucial. \par
On the other hand, for the increasing numbers of these intelligent applications, including remote monitoring, control, and inference over networks, the scalability of the network architecture requires a paradigm shift in the communication system and protocol design. This paradigm shift is for networks to focus on the effective accomplishment of the sensing task at the destination side rather than be a vehicle to solve the transmission problem \cite{shannon1949mathematical} whose goal is to reliably transmit the data produced by a source. Solving the effective communication problem in remote intelligence with efficient use of network resources requires combining the data generation and transmission problems whereby data samples that are most significant to the computation at the destination side are delivered in a timely manner\cite{uysal2022semantic}. We will refer to this paradigm as \textit{Goal-oriented} communication. Recent activities in the communication and control communities illustrate the value of such an approach in reducing communication requirements for a given level of application performance \cite{voortman2023remote, sagduyu2023multi, merluzzi2022effective, talli2023semantic}. \par
A goal-oriented communication network discards the assumption that data arrives at the communication system interface as part of an exogenous process. This implies that the communication link can pick and choose which data packets to transmit to the source for effective computation at the destination based on the state of the network (e.g., the delay state). As we do not expect the link layer to operate jointly with the application layer for all possible applications, a surrogate metric that the link layer can use when making the decisions of which packets to choose and send is useful. In this paper, following the model in \cite{shisher2022does}, we use Age of Information (AoI) as an intermediate metric that can be tracked by the link, irrespective of the particular application. A mapping of application performance with respect to the AoI is then all that is needed for the link layer to operate in a goal-oriented manner. This paper builds on the work in the series of papers  \cite{shisher2022does, sun2019sampling, ornee2021sampling,shisher2021age} that first demonstrated the usefulness of AoI metric as a surrogate for effective communication for remote tracking or inference of a process. \par
AoI is an indicator of the freshness of the data at the destination of a data flow \cite{kaul2012real}. The AoI at time $t$, $\Delta(t)$, is given by the relation $\Delta(t)=t-U_{t}$, where $U_{t}$ is the generation time of the most recently delivered data packet. This formulation is perhaps most meaningful when the data packets contain status updates, such that the most recent update makes all the past ones obsolete, yet age and functions thereof have also served very well in capturing performance related to freshness in a variety of other applications, through optimizing nonlinear functions of age. Examples of the use of generalized functions of AoI include control system scenarios \cite{klugel2019aoi, soleymani2019stochastic}, remote estimation \cite{sun2019sampling, ornee2021sampling, pan2023sampling}, and remote inference \cite{shisher2021age, shisher2022does, shisher2023learning}. While a number of studies considered the analysis of age in queuing models, closest to the spirit of this paper is the control of age via replacement of exogenous data arrivals with the generation of data "at will" \cite{bacinoglu2015age, sun2017update, kadota2018scheduling, kadota2019scheduling, sun2019sampling2, tripathi2019whittle, klugel2019aoi, sun2019closed, bedewy2021optimal}. A generalization of this approach is to incorporate jointly optimal sampling and scheduling policies to control not only age but a more sophisticated end-to-end distortion criterion by using age as an auxiliary parameter \cite{sun2019sampling, ornee2021sampling,  bedewy2020optimizing, shisher2023learning, shisher2022does}. While strikingly more demanding of analysis, these formulations take us closer to goal-oriented communication system design. \par
Almost all previous studies on the ``generate-at-will'' model adopted an assumption that the penalty of information aging is a monotonic non-decreasing function of the AoI \cite{sun2019sampling, ornee2021sampling, sun2017update, kadota2018scheduling, kadota2019scheduling, sun2019sampling2, tripathi2019whittle, klugel2019aoi, sun2019closed, bedewy2021optimal, bedewy2020optimizing}. However, it was shown in \cite{shisher2021age, shisher2022does, shisher2023learning} that the monotonicity of information aging depends heavily on the divergence of the time-series data from being a Markov chain. If the input and target data sequences in a system can be closely approximated as a Markov chain, then the penalty increases as the AoI grows; otherwise, if the data sequence significantly deviates from a Markovian structure, then the monotonicity assumption does not hold. The most closely related work to this paper is \cite{shisher2022does}, which developed scheduling policies for remote inference considering the possibly non-monotonic dependency of the AoI and practical performance. The paper \cite{shisher2022does} considered random IID delay for packet transmissions from the transmitter to the receiver and assumed a delay-free feedback channel from the receiver to the transmitter. In practical scenarios, there may be a feedback delay \cite{tsai2021unifying, pan2022optimal}, and this may affect the performance of the remote inference algorithm. Moreover, in the light of the application scenarios mentioned above (e.g., satellite and space communication), the IID delay may not be a good model, requiring a model that captures the memory in the delay process as in \cite{sun2017update}. Delayed feedback and significantly varying distribution of transmission delay are the main differences of this paper from \cite{shisher2022does}, and these differences require more technical efforts to solve the problem, as explained later in Section III. To that end, the following are the technical contributions of this paper:
\begin{itemize}
    \item We developed an index-based threshold policy to minimize the steady-state time-average inference error of an intelligent application whose input data packets are transmitted from a distant location. Given the age penalty function corresponding to a particular application, not necessarily monotonic, our policy ensures optimal practical performance, making the scheduler operate in a \textit{Goal-oriented} manner.
    \item The policy in \cite{shisher2022does} cannot achieve the optimal in the systems where channel delay statistics vary significantly with memory because of IID transmission delay and immediate feedback assumption. Hence, we expanded upon the formulation introduced in \cite{shisher2022does}, incorporating the Markovian transmission delay, with a finite number of states, and removing the assumption of immediate feedback. Although this extension is crucial for eliminating the limitations of the policy in \cite{shisher2022does}, it was not clear whether such a nicely structured policy still existed. The developed policy preserves the structure and provides a solution for practical systems with significantly varying delay statistics. Such systems will be abundant due to the envisioned coexistence of terrestrial and non-terrestrial connections in 6G, as mentioned above.
    \item The simulation results highlight the significance of (i) the policy being adaptive to non-monotonic age penalty functions and (ii) considering the memory of the delay. Numerically, the generate-at-will + zero-wait policy achieves $33\%$ to $57\%$ higher inference error than the developed policy. Furthermore, the performance gain from incorporating delay memory increases by up to $13\%$ with growing delay memory.
\end{itemize}

\section{System Model and Problem Formulation}
\begin{figure}[t]
\centerline{\includegraphics[width=.5\textwidth]{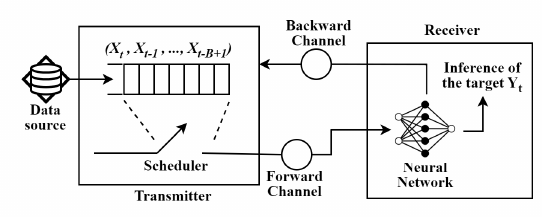}}
\caption{The remote inference system with two-way communication where the neural network on the receiver side predicts the current value of a target variable using the most recently delivered packet.}
\label{system_Model}
\end{figure}

We consider a discrete-time remote inference system, as depicted in Fig. \ref{system_Model}. The transmitter incorporates a data buffer and a scheduler. At each time slot $t$, a new data packet $X_{t} \in \mathcal{X}$ is sampled from the source and added to the buffer; meanwhile, the oldest is discarded. By this, the buffer contains the most recently sampled $B$ packets $(X_{t}, X_{t-1}, ..., X_{t-B+1})$ at any time slot $t$. The scheduler can submit any packet stored in the buffer to the channel at suitable time slots. The Age of Information (AoI), $\Delta(t)$, is the time difference between the current time $t$ and the generation time $t-\Delta(t)$ of the most recently delivered packet $X_{t-\Delta(t)}$. The value of AoI depends on (i) which packet in the buffer is selected by the scheduler during the submission of the most recently delivered packet and (ii) communication delays. A trained neural network on the receiver side takes the packet $X_{t-\Delta(t)}$ and the AoI $\Delta(t)$ as inputs and produces an action $a \in \mathcal A$ to infer the current value of a target signal $Y_{t} \in \mathcal{Y}$ at each time slot $t$. Given the AoI $\Delta(t)=\delta$, the average inference error is \cite{shisher2022does}
\begin{equation} \label{eqn:0}
    h(\delta) = \mathbb{E}_{Y, X  \sim P_{Y_{t}, X_{t-\delta}}}[L(Y,\phi(\delta, X_{t-\delta})],
\end{equation}
where $\phi:\mathbb{Z^{+}}\times\mathcal{X}\mapsto\mathcal{A}$ is the function representing the neural network, $L:\mathcal{Y}\times\mathcal{A}\mapsto\mathbb{R}$ is a loss function, and $P_{Y_{t}, X_{t-\delta}}$ is the joint distribution of the target $Y_{t}$ and the packet $X_{t-\delta}$. $L(y,a)$ is the incurred loss if the output $a$ is used for prediction when the target signal $Y_{t}=y$.\par
The loss function $L$ is determined by the purpose of the application. For example, a quadratic loss function $L_{2}(y,\hat{y})=\lVert y-\hat{y} \rVert_{2}^{2}$ is used in neural network based minimum mean-squared estimation, where the action $a=\hat{y}$ is an estimate of the target $Y_{t}=y$ and $\lVert y \rVert_{2}$ is the Euclidean norm. In softmax regression (i.e., neural network based maximum likelihood classification), the action $a=\mathcal{Q}_{Y}$ is a distribution of $Y_{t}$, and the loss function $L_{log}(y,\mathcal{Q}_{Y})=-log\mathcal{Q}_{Y}(y)$ is the negative log-likelihood function of the value $Y_{t}=y$. \par
The communication is carried out between the transmitter and receiver over a two-way channel. The two-way channel has a finite number of states $\mathcal{C} = (1, 2, ..., C)$. Each state $c$ corresponds to a distinct transmission and feedback delay distribution represented by the discrete random variables $Q_{c}$ and $R_{c}$, respectively, where $\mathbb{E}[Q_{c}]<\infty \text{ }$ and $\mathbb{E}[R_{c}]<\infty$ for all $c$. The delay duration is at least one time slot for each packet or feedback transmission. In practice, the state $c$ of the channel could be used to characterize time-varying physical phenomena that affect the transmission and feedback delay distribution, $Q_c$ and $R_c$. Examples of such phenomena include (i) the transmission duration varying due to an erasure channel situation leading to possibly a number of retransmissions, (ii) long or short network routes that packets may go through, and (iii) the delay increasing or decreasing with network congestion varying in time. Upon the delivery of the $i$-th packet, the receiver is able to detect the channel state $c_{i}$ valid during the transmission of packet $i$. Then, the channel state $c_{i}$ is fed back to the transmitter along with an ACK. The two-way channel is reliable; that is, no packet or message is lost during the transmission.\par
Recent studies \cite{shisher2022does, ShisherTimely} have exhibited scenarios where the inference error in \eqref{eqn:0} is not monotonic with age, where fresher packets do not always lead to a smaller inference error than stale ones. In other words, in certain scenarios, an older packet with AoI $>0$ can be more relevant than a fresher packet, in the sense that this packet can lead to a smaller inference error than a packet with {AoI~ $= 0$} \cite{shisher2022does, ShisherTimely}. In these cases, the "generate-at-will" approach, which samples a fresh packet upon making a transmission decision, will not be general enough to achieve optimal performance. In recognition of this fact, the "selection-from-buffer" model was proposed in \cite{shisher2022does, shisher2023learning}. In this paper, we adopt this latter model. \par
The system starts to operate at time slot $t=0$. We assume that the buffer initially contains the packets $(X_{0}, X_{-1}, ..., X_{-B+1})$. By this, the buffer is kept full at all time slots $t\geq 0$. At time slot $S_i$, the packet $X_{S_i-b_i}$ is submitted to the channel, which is the $(b_i + 1)$-th freshest packet in the buffer, with $b_i \in \mathcal{B}=\{0, 1, \ldots, B-1\}$. The packet is delivered to the receiver at time slot $D_{i}$, and the transmitter receives the corresponding ACK and channel state $c_i$ at time slot $A_{i}$, such that $S_i < D_i < A_i$. We assume that the transmitter always waits for the ACK before submitting a new packet to the channel. In other words, the scheduler always remains silent at the time slots between $S_{i}$ and $A_{i}$ for all $i$. Let $i$-th epoch be composed of the time slots in the interval $[A_{i-1}, A_{i})$. The channel state $c_{i}$ specifies the transmission delay distribution in the $i$-th epoch and evolves following a finite-state ergodic Markov chain with the transition probabilities $p_{ij}$, where $i,j\in \mathcal{C}$, and $p_{ij}$ is the transition probability from state $i$ to state $j$. The Markov chain makes a single transition at the time slots $A_{i}$ and none otherwise. Let $T_{i}=D_{i}-S_{i}$ and $F_{i}=A_{i}-D_{i}$ be the incurred transmission and feedback delays in $i$-th epoch, respectively. The distributions of $T_{i}$ and $F_{i}$ are determined by the channel state in the $i$-th epoch. That is, $p_{T_i|(c_i=c)}(q)=p_{Q_c}(q)$ and $p_{F_i|(c_i=c)}(r)=p_{R_c}(r)$, where $p_{K}(k)$ is the probability mass function (PMF) of the discrete random variable $K$. The AoI at time slot $t$ is given by
\begin{equation} \label{eqn:1}
    \Delta(t) = t-S_{i}+b_{i}, \text{  if  } D_{i} \leq t < D_{i+1}. 
\end{equation}
The initial conditions of the system are $S_{0}=0$, $\Delta(0)<\infty$, $c_{0} \in \mathcal{C}$, and $b_0 \in \mathcal{B}$. \par
We strive to minimize the expected inference error per time slot in steady-state. Accordingly, the scheduler on the transmitter side has to determine (i) when to submit a packet to the channel and (ii) which packet in the buffer to submit. A scheduling policy is a tuple $(f, g)$, where the buffer position sequence $f=(b_{1}, b_{2}, ...)$ controls which packet in the buffer to submit in each epoch, and the packet submission time sequence $g=(S_{1}, S_{2}, ...)$ specifies when to submit the packet to the channel and start the transmission in each epoch. Let $\Pi$ be the set of all policy tuples $(f, g)$. Then, the problem formulation is given by
\begin{equation} \label{eqn:2}
    h_{opt}=\inf_{(f,g)\in \Pi}\limsup_{T\to\infty} \frac{1}{T}\mathbb{E}_{(f,g)}\Bigg[\sum_{t=0}^{T-1} h(\Delta(t))\Bigg],
\end{equation}
where $h(\Delta(t))$ is the inference error at time slot $t$, and $h_{opt}$ is the optimum value of \eqref{eqn:2}. \par
This problem formulation extends the single-source scheduling problem in \cite{shisher2022does} to more practical systems by considering the memory of the transmission delay and by removing the delay-free feedback assumption.

\section{Optimal Scheduling Policy}
We provide an optimal scheduling policy to (\ref{eqn:2}) in two steps. First, we consider a class of policy tuples $(f_{\psi},g) \in \Pi$ such that $f_\psi = (b_1,b_2,\ldots)$ is a stationary deterministic buffer position sequence, where
\begin{equation} \label{eqn:3}
    b_{i+1} = \psi(c_{i}),
\end{equation} for all $i = \{0,1,2,\ldots\}$, and $\Psi$ is the set of all mapping functions $\psi: \mathcal C \mapsto \mathcal B$. Given a stationary deterministic buffer position sequence $f_\psi = (\psi(c_0), \psi(c_1), \ldots ) $, we first optimize the packet submission time sequence $g = (S_1,S_2,\ldots)$: 
\begin{equation} \label{eqn:4}
    h_{\psi,opt}=\inf_{(f_{\psi},g)\in \Pi}\limsup_{T\to\infty} \frac{1}{T}\mathbb{E}_{(f_{\psi},g)}\Bigg[\sum_{t=0}^{T-1} h(\Delta(t))\Bigg],
\end{equation}
where $h_{\psi,opt}$ is the optimum value of \eqref{eqn:4}. In the second step, we present an optimal scheduling policy for the original problem \eqref{eqn:2} using the solution to \eqref{eqn:4}. \par
The optimal packet submission time sequence $g$ for \eqref{eqn:4} will be described by using an index function defined by
\begin{equation} \label{eqn:8}
    \gamma(\delta, c)=\inf_{\nu\in\mathbb{Z}^{+}} \frac{1}{\nu}\sum_{k=0}^{\nu-1}\mathbb{E}\bigg[h(\delta+T_{i+1}+k)\bigg|c_{i}=c\bigg],
\end{equation}
where $\delta \in \mathbb{Z}^{+}$ and $c \in \mathcal{C}$.
\begin{theorem} \label{th:1}
    If $|h(\delta)|<M$ for all $\delta = 1,2,\ldots$, then the  packet submission time sequence $g=(S_{1}(\beta_{\psi}), S_{2}(\beta_{\psi}), ...)$ is an optimal solution to \eqref{eqn:4}, where
    \begin{equation} \label{eqn:10}
        S_{i+1}(\beta_{\psi}) = \min\{t\geq A_{i}(\beta_{\psi}):\gamma(\Delta(t), c_{i}) \geq \beta_{\psi}\},
    \end{equation} and $\beta_{\psi}$ is the unique root of
    \begin{equation} \label{eqn:11}
        \mathbb{E}\Bigg[\sum_{t=A_{i}(\beta_{\psi})}^{A_{i+1}(\beta_{\psi})-1} h(\Delta(t))\Bigg]-\beta_{\psi}\mathbb{E}[A_{i+1}(\beta_{\psi})-A_{i}(\beta_{\psi})] =0,
    \end{equation}
    where $A_{i}(\beta_{\psi})=S_{i}(\beta_{\psi})+T_i+F_i$ is the $i$-th ACK feedback time and $\Delta(t)=t-S_{i}(\beta_{\psi})+b_i$ is the AoI at time slot $t$. Moreover, $\beta_{\psi}$ is exactly the optimum value of (\ref{eqn:4}), i.e., $\beta_{\psi}=h_{\psi,opt}$.
\end{theorem}
\begin{proof}[Proof sketch]
The problem \eqref{eqn:4} is an infinite-horizon average-cost semi-Markov decision process (SMDP) \cite[Chapter 5.6]{bertsekasdynamic}.  Define $\tau=S_{i+1}-A_i$ as the waiting time for sending the $(i+1)$-th packet after the ACK feedback of the $i$-th packet is delivered to the transmitter. Given $\Delta(A_i)=\delta$ and $c_i=c$, the Bellman optimality equation of the  average-cost SMDP \eqref{eqn:4} is 
\begin{align}\label{relativeValuederived}
&V_{\psi}(\delta, c)=\nonumber\\
&\inf_{\tau \in \{0, 1, \ldots\}}~\mathbb E \left [ \sum_{k=0}^{\tau+T_{i+1}-1} \bigg(h(\delta+k) -h_{\psi,opt}\bigg) \bigg| c_i=c\right]\nonumber\\
&+\mathbb E \left [ \sum_{j=0}^{F_{i+1}-1} \bigg(h(\psi(c_i)+T_{i+1}+j) -h_{\psi,opt}\bigg) \bigg| c_i=c\right]\nonumber\\
&+\mathbb E[V_{\psi}(\psi(c_i)+T_{i+1}+F_{i+1}, c_{i+1})| c_i=c], 
\end{align}
for all $\delta \in \mathbb{Z}^{+}$, $\psi(\cdot) \in \Psi$, and $c \in \mathcal C$, where $V_{\psi}(\cdot)$ is the relative value function of the average-cost SMDP \eqref{eqn:4}.
Theorem \ref{th:1} is proven by directly solving the Bellman optimality equation \eqref{relativeValuederived}. Due to space limitation, the detailed proof of Theorem 1 is relegated to our technical report \cite{technical_report}.
\end{proof}
Theorem \ref{th:1} signifies that the optimal solution to \eqref{eqn:4} is an index-based threshold policy, where the index function depends on both the AoI and the channel state in the previous epoch. The scheduler submits $(i+1)$-th packet according to \eqref{eqn:3} at the first time slot for which the following conditions are satisfied: (i) the ACK of the previous transmission is received, i.e., $t\geq A_{i}(\beta_{\psi})$, and (ii) the index function exceeds the threshold. The threshold $\beta_{\psi}$ is exactly the optimum value of \eqref{eqn:4} and can be found by solving equation \eqref{eqn:11} using the low-complexity algorithms, e.g., bisection search, provided in \cite[Algorithms~1-3]{ornee2021sampling}. The index function in \eqref{eqn:8} is obtained by directly solving the Bellman optimality equation \eqref{relativeValuederived}. \par
Recent studies \cite{shisher2022does, shisher2023learning} have revealed a connection between the AoI optimization problems, such as \eqref{eqn:4}, and the \emph{restart-in-state} formulation of the Gittins index in \cite[Ch.~2.6.4]{gittins2011multi}. The optimal scheduling policy in Theorem \ref{th:1} has been obtained by employing that connection. \par
Now, we present an optimal solution to problem \eqref{eqn:2}.
\begin{theorem} \label{Th:2}
    If $|h(\delta)|<M$ for all $\delta = 1,2,\ldots$, then the policy tuple $(f^*,g^*)$ is an optimal solution to \eqref{eqn:2} such that
    \begin{itemize}
        \item[i.)] $f^*=(b^*_{1}, b^*_{2}, \ldots)$, where
        \begin{align}
            b^*_{i+1} &= \psi^*(c_{i}) \text{, } i=0,1,2,\ldots, \\
            \psi^* &= arg\min_{\psi \in \Psi} h_{\psi,opt},
        \end{align} and $h_{\psi,opt}$ is the optimum value of \eqref{eqn:4}.
        \item[ii.)] $g^*=(S^*_{1}, S^*_{2}, \ldots)$, where
        \begin{multline} \label{eqn:29}
            S^*_{i+1} = \min\{t\geq S^*_{i}+T_{i}+F_{i}:\gamma(\Delta(t), c_{i}) \geq h_{opt}\},
        \end{multline} and
        \begin{equation} \label{eqn:30}
            h_{opt}=\min_{\psi \in \Psi} h_{\psi,opt}.
        \end{equation}
    \end{itemize}
\end{theorem}
\begin{proof} 
Due to space limitation, the proof of Theorem 2 is relegated to our technical report \cite{technical_report}.
\end{proof} \par
Theorem \ref{Th:2} points out that the optimal buffer position at any submission time slot is independent of the current AoI and only depends on the channel state in the previous epoch. The reasoning behind such a result can be understood deeply by analyzing the system. At the beginning of epoch $[D_{i}, D_{i+1})$, i.e., at time slot $D_{i}$, the AoI of the process resets to $b_{i}+T_{i}$, indicating that the buffer position $b_{i}$ determines the AoI reset value for time slot $D_{i}$. The epoch $[D_{i}, D_{i+1})$ evolves starting from this AoI value. Then, the AoI resets again at time slot $D_{i+1}$ following the delivery of the subsequent packet, and the AoI values in epoch $[D_{i}, D_{i+1})$ become irrelevant to the future evolution of the process. In other words, each epoch of the process has an isolated AoI evolution, and $b_{i}$ is the parameter that allows us to control AoI evolution in $i$-th epoch. Apart from $b_{i}$, $T_{i}$ is the other parameter affecting the AoI reset value for time slot $D_{i}$, and $c_{i-1}$ reveals the distribution of $T_{i}$ thanks to the Markovian structure. Thus, $b_{i}$ is chosen, considering the distribution of $T_{i}$, through $c_{i-1}$, to ensure the AoI of the process reset to the desired value at time slot $D_{i}$. Due to this inherent structure of the process, as shown in Theorem \ref{Th:2}, the optimal buffer position sequence to problem \eqref{eqn:2} is $f^*=f_{\psi^*}=(\psi^*(c_0), \psi^*(c_1), \ldots )$, where $\psi^* \in \Psi$ is the optimal mapping function. Since $\psi^* \in \Psi$, the optimal packet submission time sequence is obtained by Theorem \ref{th:1}. \par
Even though the optimal policies outlined in Theorems \ref{th:1} and \ref{Th:2} maintain the nice structure of the policies in \cite{shisher2022does}, there are two critical distinctions. Firstly, the derived index function for this problem formulation is not solely dependent on the AoI but also incorporates the channel state. The existence of such an index function was not evident in the presence of a finite number of channel states, necessitating additional technical efforts for derivation. Secondly, in contrast to \cite{shisher2022does}, where the optimal buffer position sequence is constant due to the assumption of IID transmission delay and immediate feedback, this problem formulation considers transmission delay and non-zero feedback delay significantly varying with memory. Consequently, achieving optimal performance requires a buffer position sequence that can adapt to these variations in delay statistics. This paper unveils the relationship between the buffer position sequence and delay memory while developing Theorems \ref{th:1} and \ref{Th:2}. \par
In the special case that the inference error $h(\delta)$ is a non-decreasing function of the AoI $\delta$, the index function $\gamma(\delta,c)$ in \eqref{eqn:8} reduces to $\mathbb{E}[h(\delta+T_{i+1})|c_{i}=c]$. Moreover, the optimal buffer position sequence is $f^*=(0,0,\ldots)$.

\begin{figure}[t]
\centerline{\includegraphics[width=.43\textwidth]{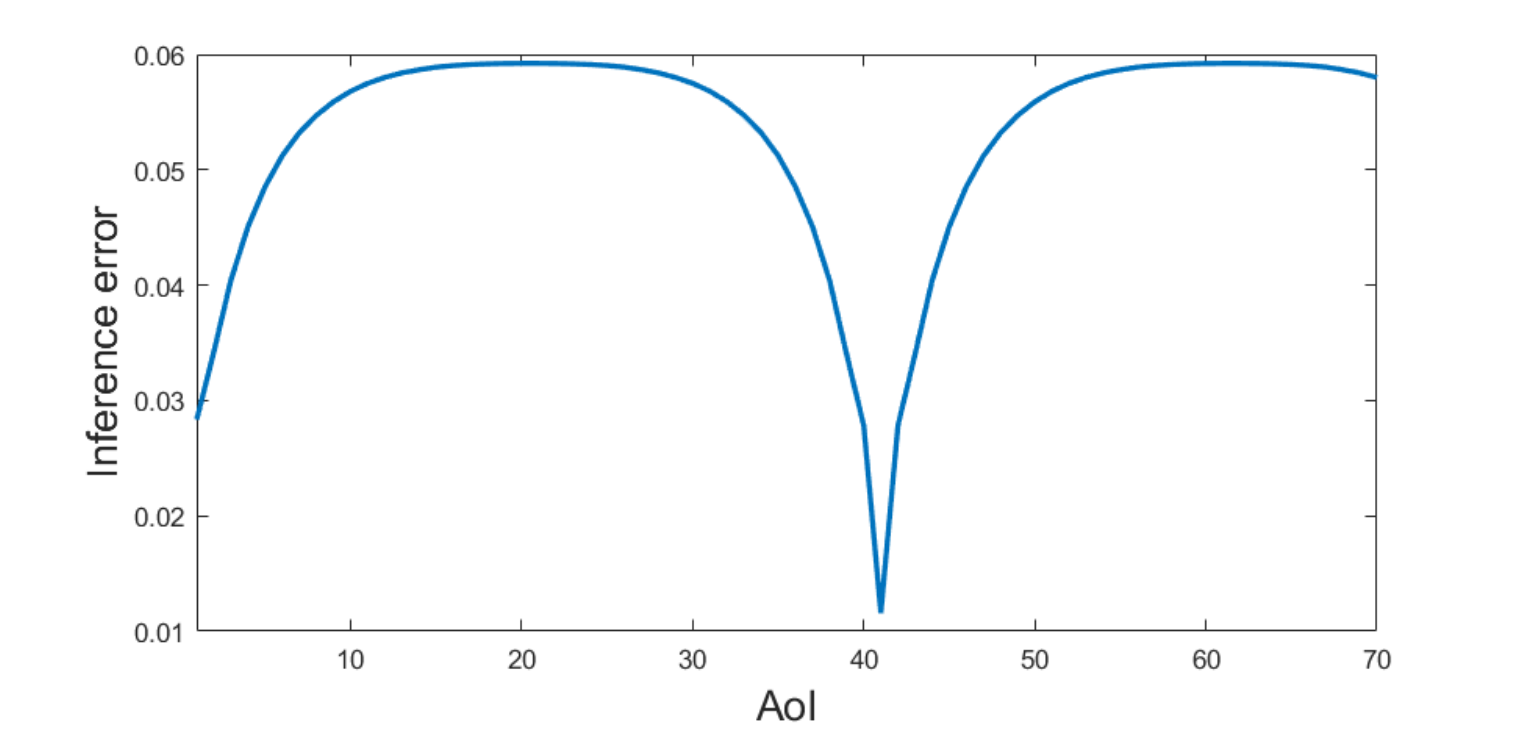}}
\caption{The incurred inference error when the current sample of the target $Y_{t}$ is predicted using a past sample $X_{t-\delta}$ with AoI $\delta$ ranging from 1 to 70.}
\label{IEvsAoI}
\end{figure}

\section{Simulation Results}
In this section, we demonstrate the performance of our optimal scheduling policy in Theorem \ref{Th:2}. To obtain an inference error function $h(\delta)$, we use a discrete-time autoregressive model of the order $p$ (AR($p$)):
\begin{align}
X_{t}=a_1 X_{t-1}+ a_2 X_{t-2}+\ldots+a_p X_{t-p}+W_t,
\end{align}
where the noise $W_t \in \mathbb R$ is zero-mean Gaussian with variance $\sigma_W^2$ and $X_t \in \mathbb R$. Let $Y_t=X_t+N_t$ be the target variable, where the noise $N_t$ is zero-mean Gaussian with variance $\sigma_N^2$. Fig. \ref{IEvsAoI} indicates the average inference error values incurred when the target $Y_t$ is predicted using $X_{t-\delta}$, where the value of AoI $\delta$ ranges from 1 to 70. We set $\sigma_W^2=0.01$, $\sigma_N^2=0.001$ and construct an AR($50$) with coefficients $a_{38}=a_{44}=0.007$, $a_{39}=a_{43}=0.05$, $a_{40}=a_{42}=0.1$, and $a_{41}=0.68$. The remaining coefficients are zero. We consider a quadratic loss function $L_{2}(y,\hat{y})=\lVert y-\hat{y} \rVert_{2}^{2}$. Because $X_{t-\delta}$ and $Y_t$ are jointly Gaussian, and the loss function is quadratic, the optimal inference error performance is achieved by a linear MMSE estimator. Hence, a linear regression algorithm is adopted in our simulations. However, our study can be readily applied to other neural network-based predictors.\par
We compare the performance of three scheduling policies: (i) the optimal policy in Theorem \ref{Th:2}, (ii) the single-source scheduling policy in \cite{shisher2022does}, and (iii) generate-at-will + zero wait policy: $(f,g)$ is such that $f=(0, 0, \ldots)$ and $g=(A_{1}, A_{2}, \ldots)$. \par
The simulation parameters related to the channel statistics have been set as follows: (i) the number of channel states $C=2$, (ii) the random variable $Q_{1}$ has the PMF $p_{Q_{1}}(3)=0.7$, and $p_{Q_{1}}(4)=0.3$, (iii) the random variable $Q_{2}$ has the PMF $p_{Q_{2}}(11)=0.35$, and $p_{Q_{2}}(12)=0.65$, (iv) the random variable $R_{1}$ has the PMF $p_{R_{1}}(1)=1$, and (v) the random variable $R_{2}$ has the PMF $p_{R_{2}}(4)=1$. The memory of the transmission delay is adjusted by varying the parameter $\alpha=p_{12}+p_{21}$ between $0$ and $2$. We set $p_{12}=p_{21}$, which ensures that the fraction of epochs with convenient delay condition is always $0.5$. The scheduling policy in \cite{shisher2022does} assumes both channel states are equally likely at each epoch independent of the history for any value of $\alpha$.\par
\begin{figure}[t]
\centerline{\includegraphics[width=.44\textwidth]{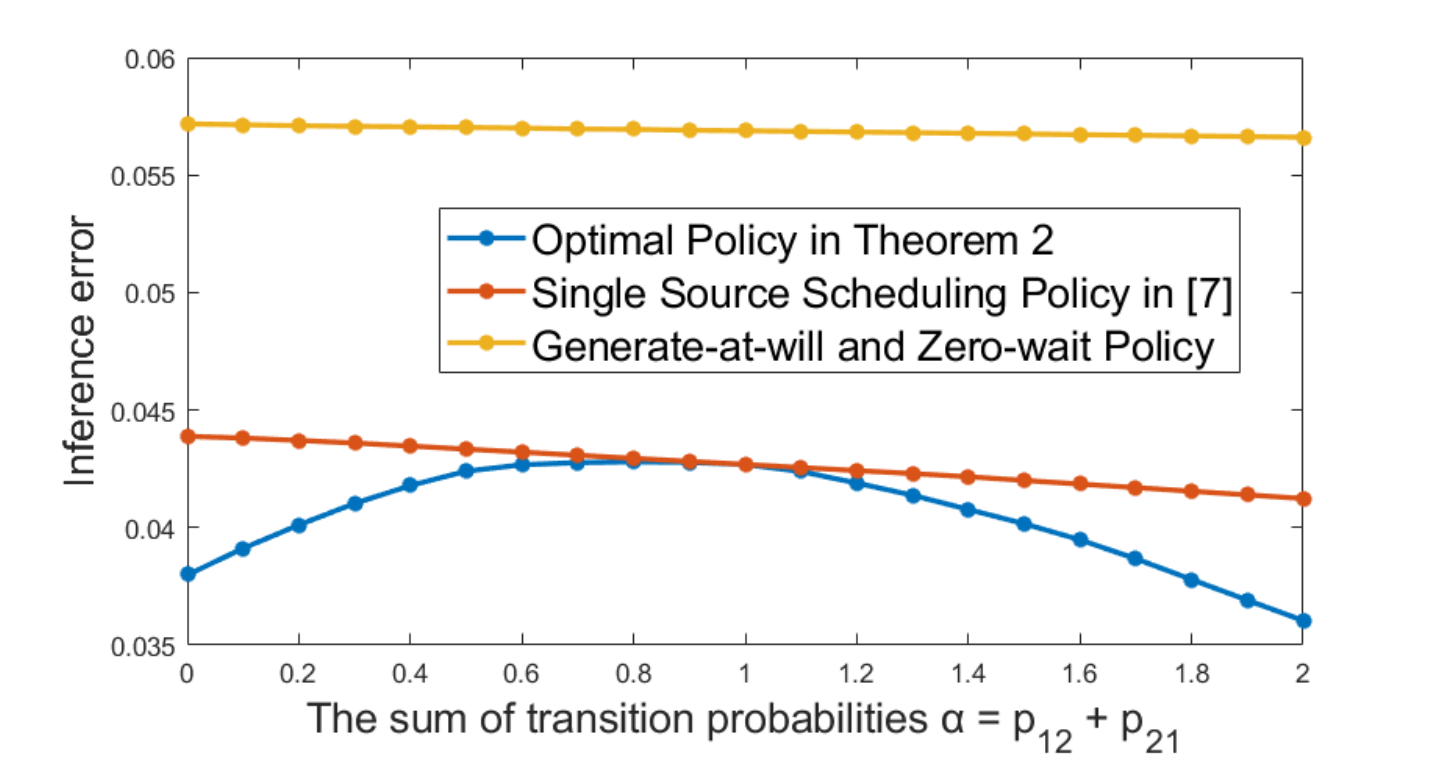}}
\caption{The time-average inference error achieved by the three different scheduling policies.}
\label{final_Result}
\end{figure}

Fig. \ref{final_Result} presents the time-average inference error values achieved by the three scheduling policies mentioned above. The results underscore that consistently submitting the freshest packet with zero waiting time leads to a $33\%$ to $57\%$ higher inference error compared to the optimal policy outlined in Theorem \ref{Th:2}. Additionally,
the optimal policy shows minimal improvement compared to the corresponding policy in \cite{shisher2022does} for $\alpha$ values around $1$. The reason for the negligible improvement is that the delay distribution is exactly IID when $\alpha=1$. However, as $\alpha$ deviates from this point, resulting in increased delay memory, the advantage of the optimal policy described in Theorem \ref{Th:2} becomes apparent, leading to an important performance gain of up to $13\%$.

\section{Conclusion}
In this paper, we studied a remote inference problem where a neural network on the receiver side predicts the real-time value of a target signal using the data packets transmitted from a distant location. Motivated by the foreseen coexistence of terrestrial and non-terrestrial connections in 6G, we assumed it would be typical for a data flow to be served through multiple alternative paths and considered transmission and feedback delay distributions significantly varying with memory. For this system model, we developed an optimal index-based threshold policy that minimizes the expected inference error per time slot in steady-state. The policy optimizes the performance for a given inference error function specific to the application and goal on the receiver side, addressing the expectation for future networks to operate in a goal-oriented manner. Finally, we demonstrated, with the simulation results, the performance gain that can be achieved in the remote inference problems by considering the memory of the delay. \par

\newpage

\bibliographystyle{IEEEtran}
\bibliography{ISIT}